\documentclass[12pt,preprint]{aastex}
\usepackage{graphicx}
\begin{document}
\title{Electron-ion coupling upstream of relativistic collisionless shocks}

\author{Yuri Lyubarsky}
 \affil{Physics Department, Ben-Gurion University, P.O.B. 653, Beer-Sheva 84105, Israel}

\begin{abstract}
It is argued and demonstrated by particle-in-cell simulations that
the synchrotron maser instability could develop at the front of a
relativistic, magnetized shock. The instability generates strong
low-frequency electromagnetic waves propagating both upstream and
downstream of the shock. Upstream of the shock, these waves make
electrons lag behind ions so that a longitudinal electric field
arises and the electrons are accelerated up to the ion kinetic
energy. Then thermalization at the shock front results in a plasma
with equal temperatures of electrons and ions. Downstream of the
shock, the amplitude of the maser-generated wave may exceed the
strength of the shock-compressed background magnetic field. In
this case the shock-accelerated particles radiate via nonlinear
Compton scattering rather than via a synchrotron mechanism. The
spectrum of the radiation differs, in the low-frequency band, from
that of the synchrotron radiation, providing possible
observational tests of the model.
\end{abstract}


\maketitle

\section{Introduction.}
Relativistic shocks are supposed to play an important role in
various astrophysical objects, e.g., in pulsar wind nebulae,
active galactic nuclei, gamma-ray bursts. In all these cases
shocks are collisionless
therefore some plasma instabilities are assumed to provide
dissipation necessary for the final thermalization of the flow. An
important point is that in relativistic case, plasma instabilities
are generally electromagnetic and therefore strong low-frequency
electromagnetic waves could be emitted. If a non-negligible
magnetic field is presented in the upstream flow, the synchrotron
maser instability develops at the shock front (Langdon, Arons \&
Max 1988).

One-dimensional particle-in-cell (PIC) simulations of relativistic
magnetized shocks in electron-positron and electron-positron-ion
plasmas (Gallant et al. 1992; Hoshino et al. 1992) demonstrate
intensive electromagnetic waves both upstream (precursor) and
downstream of the shock; these waves are generated by synchrotron
maser instability at the shock front. It was shown that the
precursor takes a few-per cent of the flow energy. One can
anticipate that the upstream flow could be significantly affected
by interaction with the precursor; this in turn could
significantly influence the structure of the shock. For example,
absorption of the precursor would result in strong deceleration
and heating of the flow even though the energy of the precursor is
small compared with the total energy of the flow; this follows
immediately from the conservation of energy and momentum in the
ultrarelativistic case. Of course true absorption is negligibly
weak in the case of interest; the precursor radiation should be
eventually absorbed via non-linear plasma processes, e.g., induced
scattering. In this case one can anticipate formation of
non-thermal particle distribution already in the upstream flow,
which could significantly affect particle acceleration process at
the shock.

Investigation of complicated nonlinear absorption of the precursor
wave is beyond the scope of the present research; this would
require multidimensional simulations at a very large scale. The
aim of the present study is to consider a simpler mechanism of
interaction between the electromagnetic precursor and the upstream
flow. This mechanism does not assume absorption; however, it
operates only in electron-ion flows. The basic idea is the
following. Because electrons interact with the waves whereas ions
do not, some velocity difference arises between the electron and
ion flows illuminated by powerful electromagnetic waves. This is
not associated with absorption; just because electrons experience
relativistic oscillations in the field of the strong wave, the
velocity of their guiding centers decreases. As ions proceed with
the initial velocity, a difference in the bulk velocities of
electron and ion fluid arises. Therefore a longitudinal electric
field is generated so that the electrons are accelerated whereas
the ions are decelerated. It will be shown below that in the
highly relativistic case, the energy equipartition is achieved
between the electrons and ions before the flow arrives at the
shock front.

The article is organized as follows. In sect. 2, the basic
characteristics of the electron motion in a strong electromagnetic
wave are outlined. The behavior of a homogeneous electron-ion flow
in a strong electromagnetic wave is studied in sect. 3. In sect.
4, the synchrotron maser instability at the shock front is
considered. PIC simulations of relativistic shocks are presented
in sect. 5. The results are discussed in sect. 6. In the Appendix,
a growth rate of the synchrotron instability of a relativistic,
narrow ring is estimated.

\section{Electron in a strong electromagnetic wave}
Let us first briefly outline motion of an electron in a strong
wave. Let the electron move along the $x$ axis in positive
direction and the wave propagate in the opposite direction. Let
the wave be monochromatic and polarized in the $y$-direction,
$E_y=E_0\sin\omega (x+t)$, $E_z=0$. The electron has two integrals
of motion (e.g. Landau \& Lifshitz 1987, pp 113-115, 119).
Invariance with respect to a shift in the $y$-direction implies
conservation of the $y$-component of the generalized momentum,
$p_y-eA_y=\it const$, where $\mathbf{p}=m_e\mathbf{u}$ is the
particle momentum, $\mathbf{A}$ the vector potential of the wave,
$\mathbf{u}$ the electron 4-velocity, $m_e$ the electron mass and
$e$ the positive quantity equal, in absolute value, to the
electron charge. The speed of light is taken to be unity
throughout this paper. Only the $y$-component of the vector
potential is non-zero in the linearly polarized wave,
\begin{equation}
A_y=(E_0/\omega)\cos\omega (x+t);\label{A}
\end{equation}
therefore one can write
\begin{equation}
u_y=a\cos\omega(x+t); \label{u_x}
\end{equation}
where
\begin{equation}
a\equiv \frac{eE_0}{m_e\omega} \label{strength}
\end{equation}
is the strength parameter of the wave. If $a\ll 1$, the wave is
linear, at $a>1$ oscillations of the electron in the field of the
wave become relativistic.

The second integral of motion, which follows from the equations of
motions, is $\gamma+u_x=\it const=\gamma_0(1+v_{x0})$, where
$\gamma=\sqrt{1+u_x^2+u_y^2}$ is the electron Lorentz factor. This
yields, together with Eq.(\ref{u_x}),
\begin{equation}
\gamma=\gamma_0+\frac{a^2\cos^2(x+t)}{2(1+v_0)\gamma_0}.\label{energy}
\end{equation}
In what follows, we assume
\begin{equation}
1\ll a\ll\gamma_0\label{cond}
\end{equation}
so that the wave is strong (oscillations in the guiding center
frame are relativistic) but the electron energy does not vary much
in the course of oscillation. Making use of Eqs.(\ref{u_x},
\ref{energy}) one can easily find the velocity of the electron
guiding center $v_{gc}\equiv\langle u_x/\gamma\rangle$, where the
angular brackets mean averaging over the wave period. As all the
velocities are close to the speed of light, one can conveniently
use the Lorentz factor of the guiding center frame:
\begin{equation}
\gamma_{gc}\equiv
(1-v^2_{gc})^{-1/2}=\frac{\sqrt{2}\gamma_0}{a}.\label{guidcentr}
\end{equation}
One can see that in the strong wave, the velocity of the electron
guiding center decreases. Consequently, if an electron-ion flow is
illuminated by a strong electromagnetic wave, the electrons lag
behind the ions. In this case a longitudinal electric field arises
and electrons will be accelerated whereas ions will be
decelerated.

\section{Energy exchange between electrons and ions in a strong
electromagnetic wave.} To gain an impression of how the
relativistic plasma flow interacts with a strong wave, let us
consider evolution of a spatially homogeneous relativistic plasma
flow in a wave of a constant amplitude satisfying the condition
(\ref{cond}). Let the wave be linearly polarized and propagate
towards the plasma flow.
\subsection{Non-magnetized flow.}
The electron equations of motion can be written as
\begin{equation}
m_e\frac{du_y}{dt}=e\left(\frac{\partial A_y}{\partial
t}+v_x\frac{\partial A_y}{\partial x}\right);\quad
m_e\frac{du_x}{dt}=-e\left(E_{\|} +v_y\frac{\partial A_y}{\partial
x}\right).\label{eqmotion}
\end{equation}
Here $A_y$ is the vector potential of the wave (\ref{A}), $E_{\|}$
the longitudinal (along the $x$ axis) electric field. The
longitudinal electric field arises because of mismatch in
velocities of electron and ions; it satisfies the equation
\begin{equation}
\frac{\partial E_{\|}}{\partial t}+4\pi
Ne\left(V_x-v_x\right)=0,\label{Efield}
\end{equation}
where $N$ is the particle density, $\mathbf{V}$ the ion velocity.
The last is found from the equations of motion of ions, which are
written similarly to Eqs.(\ref{eqmotion})
\begin{equation}
m_i\frac{dU_y}{dt}=-e\left(\frac{\partial A_y}{\partial
t}+V_x\frac{\partial A_y}{\partial x}\right);\quad
m_i\frac{dU_x}{dt}=e\left(E_{\|}+V_y\frac{\partial A_y}{\partial
x}\right).\label{ions}
\end{equation}
Here $m_i$ is the ion mass, $\mathbf{U}=\mathbf{V}\Gamma$ the ion
4-velocity, $\Gamma$ the ion Lorentz factor.

It follows immediately from the first Eq.(\ref{eqmotion}) that
Eq.(\ref{u_x}) remains valid in this case.    Combining
Eqs.(\ref{eqmotion}) yields the equation
\begin{equation}
m_e\gamma\frac{dS}{dt}=-eE_{\|}S;\label{ev}
\end{equation}
where $S\equiv (u_x+\gamma)/2$. Thus $u_x+\gamma$ is not constant
in the presence of a longitudinal field however it varies slowly
as compared with the wave period and therefore one can average
Eq.(\ref{ev}) over the wave period to get
\begin{equation}
m_e\frac{d\langle\gamma\rangle}{dt}=-e\langle
E_{\|}\rangle.\label{evol}
\end{equation}
Here it is taken into account that the right inequality in the
condition (\ref{cond}) leads to the relation
$S\approx\langle\gamma\rangle$ (see Eq.(\ref{energy}).

Similarly the ion equations of motion (\ref{ions}) are reduced to
\begin{equation}
U_y=-a\frac{m_e}{m_i}\cos\omega(x+t); \label{U_x}
\end{equation}
\begin{equation}
m_i\frac{d\langle\Gamma\rangle}{dt}=e\langle
E_{\|}\rangle;\label{evol1}
\end{equation}
which are equivalent to Eqs.(\ref{u_x}) and (\ref{evol}).

 Averaging Eq.(\ref{Efield}), one can see that $\langle
E_{\|}\rangle$ is determined by the velocities of the particle
guiding centers. Making use of the identity
$\gamma^2=1+u_x^2+u_y^2$, which is conveniently written as
$2S(\gamma-u_x)=1+u_y^2$, and Eq.(\ref{u_x}), one can write (cp.
Eq.({\ref{guidcentr}))
\begin{equation}
\langle
v_x\rangle=1-\langle\frac{1+u_y^2}{2S\gamma}\rangle=1-\frac{a^2}{4\langle\gamma\rangle^2}.\label{eguid}
\end{equation}
It was taken into account that under the condition (\ref{cond}),
$S$ is close to $\gamma$ and both are close to
$\langle\gamma\rangle$. Analogously one gets for the velocity of
the ion guiding center
\begin{equation}
\langle
V_x\rangle=1-\frac{2+(am_e/m_i)^2}{4\langle\gamma\rangle^2}.\label{iguid}
\end{equation}
Now one can write Eqs.(\ref{Efield}), (\ref{evol}) and
(\ref{evol1}) in the dimensionless form:
\begin{equation}
\frac{dw}{d\tau}+\frac{a^2}{4\langle\gamma\rangle^2}-\frac
{2+(am_e/m_i)^2}{4\langle\Gamma\rangle^2}=0;\label{w}
\end{equation}\begin{equation}
\frac{d\langle\gamma\rangle}{d\tau}+w=0;\label{S}
\end{equation}
\begin{equation}
\frac{d\langle\Gamma\rangle}{d\tau}-\frac{m_e}{m_i}w=0.\label{Gamma}
\end{equation}
Here $\tau=\omega_pt$, $\omega_p=(4\pi e^2N/m_e)^{1/2}$,
$w=\langle E_{\|}\rangle (4\pi nm_e)^{-1/2}$.
 It follows from Eqs. (\ref{S}) and (\ref{Gamma}) that
$m_i\langle\Gamma\rangle+m_e\langle\gamma\rangle=(m_i+m_e)\Gamma_0,$
where $\Gamma_0$ is the initial flow Lorentz factor. Then
eliminating $w$ from Eqs.(\ref{w}) and (\ref{S}) yields
\begin{equation}
\frac{d^2\langle\gamma\rangle}{d\tau^2}-\frac{a^2}{4\langle\gamma\rangle^2}+
\frac{2m_i^2+a^2m_e^2}{4[(m_i+m_e)\Gamma_0-m_e\langle\gamma\rangle]^2}=0.
\end{equation}
This equation describes nonlinear oscillations of the electron-ion
plasma. The equilibrium occurs at
\begin{equation}
\gamma_{eq}=\frac{a\Gamma_0}{\sqrt{2+(am_e/m_i)^2}+am_e/m_i}=\left\{\begin{array}{ll}
2^{-1/2}a\Gamma_0; & a\ll m_i/m_e; \\
(m_i/2m_e)\Gamma_0; & a\gg m_i/m_e.\end{array}\right.\label{equil}
\end{equation}
when the  electron and ion guiding centers move with the same
velocity (see Eq.(\ref{guidcentr})). Writing down the first
integral
\begin{equation}
\left(\frac{d\langle\gamma\rangle}{d\tau}\right)^2+
\frac{a^2}{2\langle\gamma\rangle}+\frac{2m_i^2+a^2m_e^2}{2m_e[(m_i+m_e)\Gamma_0-m_e\langle\gamma\rangle]}=
\frac{2m_i+m_ea^2}{2m_e\Gamma_0},
\end{equation}
one can see that the electron Lorentz factor oscillates around
this equilibrium from the initial value $\Gamma_0$ up to
$\min(a^2,M/m)\Gamma_0$. So if $a>\sqrt{M/m}$, the energy is
efficiently transferred from ions to electrons.
The characteristic time of the energy exchange, $T$, is estimated
as the reverse frequency of oscillations
\begin{equation}
\frac 1T=\left\{\begin{array}{ll}a^{-1/2}\Gamma_0^{-3/2}\omega_p; & a\ll m_i/m_e; \\
a(m_e/m_i\Gamma_0)^{-3/2}\omega_p; & a\gg
m_i/m_e.\end{array}\right.\label{tau}
\end{equation}
If the wave is switched on slowly, so that $a$ grows on a
time-scale large compared with $T$, the electron remains in
equilibrium and moves with the Lorentz-factor of Eq.(\ref{equil}).
Then the electron energy grows linearly with $a$ until equilibrium
with ions is achieved at $a\sim m_i/m_e$.

\subsection{Magnetized flow}
The mechanism outlined works also in magnetized flows. Note that
even if the magnetic field is tangled in the proper plasma frame,
the transverse magnetic field dominates in the lab frame therefore
it will suffice to consider the flow with the magnetic field
$B_0\mathbf{\widehat{z}}$. Initially the magnetic field is frozen
into the plasma therefore the electric field $v_0B_0{\widehat{y}}$
is presented in the lab frame. In the spatially homogeneous case,
the magnetic field remains constant; however, the electric field
varies according to
\begin{equation}
\frac{\partial E}{\partial t}=4\pi env_y.\label{E}
\end{equation}
 The equations of motion are obtained by adding the Lorentz force
\begin{equation}
\mathbf{F}=-e\gamma(u\mathbf{\widehat{y}}+\mathbf{v\times\widehat{z}})B_0;
\label{Lorentz}
\end{equation}
to the right hand side of Eqs. (\ref{eqmotion}) and the
corresponding force to Eqs.(\ref{ions}); here $u\equiv E/B_0$ is
the drift velocity. Taking into account that the wave frequency is
high, one can introduce the slowly varying variables $S\equiv
(\gamma+u_x)/2$ and $R\equiv u_y-a\cos\omega(x+t)$, which become
constant in the absence of external fields. Substituting these
variables into the electron equations of motion and averaging
yields
\begin{equation}
\frac{\omega_p}{\omega_B}\frac{dR}{d\tau}=\frac{a^2+2R^2}{4S^2}-1+u;
\label{R}
\end{equation}
\begin{equation}
\frac{\omega_p}{\omega_B}\frac{dS}{d\tau}=R+S\cal E; \label{S1}
\end{equation}
where $\omega_B\equiv eB_0/m$, ${\cal E}\equiv E_{\|}/B_0$.
Substituting the electric field $E=uB_0$ into Eq.(\ref{E}) and
averaging yields
\begin{equation}
\frac{\omega_B}{\omega_p}\frac{du}{d\tau}=-\frac
RS.\label{field_vel}
\end{equation}
Equations (\ref{R}), (\ref{S1}) and (\ref{field_vel}) govern the
evolution of the electron flow in the presence of the magnetic
field.

Let us first neglect the electron-ion coupling and put ${\cal
E}=0$. Then Eqs.(\ref{S1}) and (\ref{field_vel}) yield
\begin{equation}
r=1+2\left(\frac{\omega_p\Gamma_0}{\omega_B}\right)^2\ln
s;\label{y}
\end{equation}
where $r\equiv 2\Gamma_0^2(1-u)$, $s\equiv S/\Gamma_0$. One can
check a posteriori that $R\ll a$. Then Eq.(\ref{R}) is reduced,
with account of Eqs.(\ref{S1}) and (\ref{y}), to
\begin{equation}
\Gamma_0\frac{d^2s}{d\tau^2}=
\frac{\omega_B^2}{2\Gamma^2_0\omega_p^2}\left(\frac{a^2}{2s^2}-1\right)-\ln
s.\label{dzdt}
\end{equation}
This equation describes nonlinear oscillations of electrons caused
by discrepancy between the velocity of electron guiding centers
and the drift velocity $E/B_0$. If the flow magnetization is not
too high,
\begin{equation}
\zeta\equiv\frac{a\omega_B}{\Gamma_0\omega_p}\ll 1;\label{magn}
\end{equation}
the electron energy remains nearly constant, $s-1\ll 1$, and
Eq.(\ref{dzdt}) reduces to a linear equation
\begin{equation}
\Gamma_0\frac{d^2(s-1)}{d\tau^2}+s-1=\frac{\zeta^2}4,\label{magnosc}
\end{equation}
which describes oscillations with the frequency
$\omega_p/\sqrt{\Gamma_0}$ around the equilibrium value
$s_0=1+\zeta^2/4$. It follows from Eq.(\ref{y}) that the
equilibrium drift velocity is $u_0=1-a^2/(4\Gamma_0^2)$, which
coinsides with the velocity of the electron guiding center (see
Eq.(\ref{guidcentr})). Therefore if the wave is switched on slowly
such that $a$ grows on a time-scale larger than
$\sqrt{\Gamma_0}/\omega_p$, the system evolves remaining in the
equilibrium $s=s_0$, $u=u_0$ so that the magnetic field remains
frozen in the electron fluid in the sense that $E/B_0$ is equal to
the velocity of the guiding-center frame. In this case the force
(\ref{Lorentz}) remains, on average, close to zero and the
evolution of the system may be described by non-magnetized
equations (\ref{w}),  (\ref{S}) and (\ref{Gamma}). So at the
condition (\ref{magn}), the energy exchange between electrons and
positrons proceeds as in the non-magnetized case. Note that the
characteristic time of the electron-ion energy exchange
(\ref{tau}) significantly exceeds the period of oscillations
described by Eq.(\ref{magnosc}); this justifies neglect of the
energy exchange in the above analysis of the interaction of the
electron flow with the magnetic field.

If the condition (\ref{magn}) is violated, Eq.(\ref{dzdt})
describes non-linear oscillations around the equilibrium at which
the zero-electric-field frame coincides with the velocity of the
guiding-center frame.
The equilibrium electron energy grows with $\zeta$ and eventually
reaches $ma\Gamma_0$ so that the velocity of the zero electric
field frame is equal to the initial velocity. Thus  the electron
energy grows in any case until the velocity of the guiding center
frame approaches the initial velocity.

\section{Synchrotron maser instability at the shock front}
Relativistic shocks are conveniently normalized by the
magnetization parameters
\begin{equation}
\sigma_s=\frac{B}{4\pi Nm_s\Gamma};
\end{equation}
where $B$ is the upstream magnetic field, $N$ the upstream plasma
number density (both are measured in the shock frame), $\Gamma$
the upstream Lorentz factor. The index $s$ refers to the plasma
species ($i$ for ions, $e$ for electrons). We consider here the
case $\sigma_i\ll 1$ when the shock is strong; the electron
magnetization may be arbitrary. Note that relativistic shocks are
generally perpendicular in the sense that the magnetic field is
directed predominantly perpendicular to the shock normal. This is
true unless the field is aligned with the shock normal to within
$1/\Gamma$ in the upstream frame.

It follows from the above consideration that the energy exchange
between the ions and electrons could occur upstream of the
relativistic shock provided some electromagnetic instability at
the shock front generates strong enough electromagnetic waves
propagating in the forward direction. Langdon et al. (1988) and
Gallant et al. (1992) demonstrated that in relativistic
electron-positron shocks, synchrotron maser instability generates
strong semicoherent electromagnetic waves both upstream and
downstream of the shock. Simulations of electron-positron-ion
plasma also revealed a strong electromagnetic precursor generated
by electrons and positrons at the shock front (Hoshino et al.
1992). Heavy ions cannot emit at the frequency exceeding the
plasma frequency. In electron-positron-ion shocks, the ion
synchrotron instability develops because low-frequency
magnetosonic waves could propagate in electron-positron plasma
(Hoshino et al. 1992) however these waves do not propagate
upstream of the shock because their group velocity is less than
the flow velocity therefore even in this case the electromagnetic
precursor is generated only by electrons and positrons. In
electron-ion plasma, only the electron synchrotron instability is
possible.

Within the shock structure, the directional motion of the
particles is converted into the rotational motion in the enhanced
magnetic field; one can expect that a ring-like structure appears
in the particle phase space. This is definitely true for ions
however the ion synchrotron instability is suppressed by
electrons. The synchrotron maser instability is possible if
electrons form a ring or a horseshoe in the phase space. It is not
evident that such distributions arise within the shock structure
because the electron Larmor radius is much smaller than the shock
width so that motion of electrons is rather complicated.
Simulations of non-relativistic shocks reveal holes in the
electron phase space (Shimada \& Hoshino 2000; McClements et al.
2001; Hoshino \& Shimada 2002; Schmits, Chapman \& Dandy 2002a,b);
these were attributed to localized electrostatic structures formed
at the non-linear stage of the Buneman instability. Note that
earlier Tokar et al. (1986) found in their simulations of the
non-relativistic electron-ion shock that an extraordinary mode
noise was propagating away from the shock; Bingham et al. (2003)
interpreted this noise as the electron cyclotron maser emission
from the shock front.

In relativistic flows, development of the electrostatic
instabilities is suppressed nevertheless one can expect that
immediately after the electrons enter the shock front, a
significant fraction of their kinetic energy is converted into the
energy of the Larmor rotation in the enhanced magnetic field
because in the proper electron frame of reference, the magnetic
field increases on a time-scale less than the proper electron
gyroperiod. The electron with the Lorentz factor $\Gamma$ "sees"
the magnetic field $B'=\sqrt{B^2-E^2_{\perp}}=B/\Gamma$, where
$E_{\perp}=vB$ is the electric field perpendicular to the flow
velocity. In the one-dimensional case, $E_{\perp}$ remains
constant and therefore $B'$ varies significantly when $B$ varies
only by a fraction $1/\Gamma^2$. In the shock frame, $B$ varies at
the scale of the ion Larmor radius, $R_i=m_i\Gamma/eB$, so the
electron "sees" strong variation of the magnetic field when it
enters the shock by a depth of only $d\sim R_i/\Gamma^2$. In the
proper electron frame, the field increases for the time
$t'=d/\gamma\sim R_i/\Gamma^3$ therefore if
$\Gamma>(m_i/m_e)^{1/3}$, the electron motion becomes
non-adiabatic when it just enters the shock. In this case one can
expect formation of a ring-like distribution. In the next section,
PIC simulations will be presented confirming this conjecture.

Maser emission from the relativistic ring occurs at the harmonics
of the relativistic gyrofrequency $\Omega_B\equiv\omega_B/\Gamma$.
The growth rate of the instability is generally a rather
complicated function of the plasma parameters. In the Appendix,
the instability of a highly relativistic, narrow ring
$1<\delta\Gamma\ll\Gamma$, is considered in two limiting cases,
when the relativistic plasma frequency $\Omega_p\equiv
\omega_p/\sqrt{\Gamma}$ is much less and much larger than
$\Omega_B$. Note that $\sigma_e=\Omega_B^2/\Omega_p^2$; therefore
one can write the corresponding limits as $\sigma_e\gg 1$ and
$\sigma_e\ll 1$. The maximal growth rate and the corresponding
frequency of the emitted waves are written in these limits as
\begin{equation}
\kappa=\left\{\begin{array}{l} 0.3\sigma_e^{-1/3}\Omega_B;  \\
0.1(\Gamma/\delta\Gamma)\sigma^{1/4}\Omega_B;
\end{array}\right.\quad
\omega=\left\{\begin{array}{ll} \sigma_e^{1/2}\Omega_p; & \sigma_e \gg 1 \\
0.5\sigma^{-1/4}\Omega_p; & \sigma_e\ll 1.\end{array}\right.
\end{equation}
One sees that unless $\sigma_e$ is too large or too small, the
maser instability develops on a time-scale of a few Larmor periods
and generates emission at frequencies of a few plasma frequencies.
One can expect that this remains true also in the intermediate
regime $\sigma_e\sim 1$.

It will be shown in the next section that the electron phase space
distribution at the front of the highly relativistic shock has a
ringlike shape and that such shocks do emit strong low frequency
radiation. The strength parameter of the emitted waves may be
estimated as follows. The emission frequency was shown to be
something larger than the plasma frequency therefore one can write
$\omega=\eta\Omega_p$, where $\eta$ is a factor of the order of a
few in a very wide range of $\sigma_e$. The amplitude of the wave
is expressed via the wave power, which may be parametrized as a
fraction $\xi< 1$ of the electron upstream energy. Then the
strength parameter (\ref{strength}) can be expressed as
\begin{equation}
a=\frac{\sqrt{2\xi}}{\eta}\Gamma_0.
\end{equation}
It follows from this estimate that the precursor from a highly
relativistic shock may be considered as a strong wave even if the
emitted fraction of the electron energy is small. Therefore one
can anticipate, according to the results of sect. 3, an efficient
energy exchange between the electrons and ions upstream of the
shock.

 \section{Energy exchange between electrons and
ions upstream of the relativistic shock.}

In order to check whether the mechanism outlined is operative at
relativistic shocks, we performed PIC simulations. We used a
one-dimensional, relativistic, electromagnetic code essentially
the same as described by Birdsall \& Langdon (1991). The
simulation is one-dimensional in space along the $x$ axis, which
coincides with the flow direction. Particle motion is restricted
to $x-y$ plane, the magnetic field is oriented in the direction
$z$. The particles are advanced in time using the relativistic
Lorentz force equation. The particle momentums, not velocities,
were used as variables in order to avoid losing accuracy when
calculating expressions such as $(1-v^2)^{-1/2}$. The longitudinal
electric field is found from the Poisson equation. The transverse
fields are evolved via
Maxwell's equations reduced to equations for the right-hand and
left-hand propagating waves, $F^{\pm}=E\pm B$. These equations are
solved along the exact vacuum characteristics, $x\pm t$, such that
vacuum waves propagate exactly with the speed of light; then even
a highly relativistic plasma flow does not generates numerical
Cerenkov emission  (Birdsall \& Langdon 1991).

At $t=0$ the computation box is filled with a homogeneous plasma
moving to the right with the Lorentz factor
$\Gamma_0=(1-V_0^2)^{-1/2}$. The thermal velocity dispersion is
negligible. The plasma initially carries a uniform magnetic field,
$B_0$, as well as the electric field $E_x=V_0B_0$. The flow moves
to the right against a wall from which the particles are
elastically reflected. The plasma is continuously injected from
the left boundary. The condition of no incoming waves was adopted
at the right boundary. In Figs. 1 and 2, the simulation results
are presented for a run with the initial Lorentz factor of the
flow $\Gamma_0=50$ and the ion-to-electron mass ratio
$m_i/m_e=200$. The flow is weakly magnetized, $\sigma_i=0.003$.
The system grid size is $204,000$, the particle density is 2 per
species per cell, the time-step, $\Delta t=\Delta x$, was chosen
such that $eB_0\Delta t/mc=0.5$. The length is measured in units
of the upstream relativistic ion Larmor radius,
$R_i=M\Gamma_0/eB_0$.  The physical system size is $8.6R_i$.

According to the simulation results, a shock arises initially at
the right boundary and then moves to the left. Fig. 1 shows the
initial stage when the shock was just formed.   One can see a
well-developed ion loop in Fig. 1a. The front of the shock occurs
when the upstream flow meets the first reflected ions ($x=8.01R_i$
at the presented snapshot). The electron flow near this point is
shown in Fig. 1b; one can see that the electrons enter the shock
non-adiabatically and that their motion becomes rotational. The
ring formed in the electron phase space is clearly seen in Fig.
1c. Just below this initial region, the phase space hole
disappears because the synchrotron instability develops and
electron distribution spreads. Waves generated by the instability
are clearly seen in Fig. 1d. The electromagnetic precursor
propagates ahead of the shock front exciting oscillations of
electrons and making the electrons lag behind the ions. A
large-scale longitudinal electric field arises, which accelerates
electrons. This acceleration is seen in Fig. 1e. The electrons
start to accelerate when the precursor reach them (at $x=6.8R_i$
in the presented snapshot). At this stage, the electron energy
gain is not large and the electrons enter the shock with an energy
still much less than the energy of ions.

In the course of time, the shock propagates farther to the left
and so does the precursor. As the electrons with a larger energy
enter the shock, the amplitude of the precursor increases. In the
wave with a larger amplitude, the electron guiding center slows
down, a larger longitudinal electric field develops and, according
to Eq.(\ref{equil}), the electrons acquire larger energy. When
these new electrons enter the shock, they emit even stronger waves
and then the energy of electrons in the upstream flow increases
even more. The ion energy decreases appropriately because the
total energy of the flow is conserved. Eventually, long-wavelength
electrostatic oscillations develop. In these oscillations, the
average energies of electrons and ions are equal. Fig.2 shows the
simulation results at $t=8.2R_i$. Here the shock front is at
$x=5R_i$. The shock is sharp, the shock width is of the order of
$R_i$. Strong high frequency oscillations of the magnetic field at
$0.5<z<5$ (see Fig.1d) represent the electromagnetic precursor.
Downstream of the shock, the plasma density and the mean magnetic
field are 3 times those upstream as they should be in a
two-dimensional relativistic gas with a ratio of specific heats
$3/2$ (note that all quantities are measured in the downstream
frame; in the shock frame, the density jump is 2).

These simulations show that the energy equipartition between
electrons and ions is achieved before the plasma enters the shock
front so that downstream of the shock, the temperatures of ions
and electrons are equal. The front of the precursor was emitted
when the electrons with the initial Lorentz factor entered the
shock; therefore the wave amplitude is initially not very large
and the electrons are accelerated slowly but the amplitude grows
when the electrons with larger energies enter the shock and
therefore the precursor amplitude grows and the electrons are
accelerated more rapidly. When the electrons achieve equipartition
with ions, the amplitude of the precursor becomes maximal. At this
stage, the precursor is taking few per-cent of the total energy of
the flow.

The electrons at the shock front emit electromagnetic waves both
forward and backward. In weakly magnetized flows,
$\sigma_i\lesssim 0.01$, the energy density of the waves
downstream of the shock exceeds that of the shock-compressed
background magnetic field.

 We ran simulations with the Lorentz factor of the
flow from 2 to 50. It was found that electrons efficiently take
energy from ions provided the flow is highly relativistic,
$\Gamma_0> 10$. In the mildly or non relativistic case, the shock
does not emit waves strong enough to make electrons lag behind
ions, so no longitudinal electric field arises.

There is no sign of nonthermal particle acceleration in these
simulations; downstream of the shock, the distribution functions
of both electrons and ions are Maxwellian. One of the reasons is
that in the one-dimensional simulations, the electric field is
orthogonal to the magnetic field. The Fermi acceleration
presumably occurs at much larger scale and anyway requires
three-dimensional turbulence. However this study shows that the
nonthermal tail in the electron spectrum would begin, provided it
forms, from the energy of the order of the upstream ion energy.

\section{Discussion}
It has been demonstrated in this paper that the maser instability
develops at the front of the ultrarelativistic shock and generates
electromagnetic waves propagating both upstream and downstream of
the shock. Interaction of the upstream plasma flow with these
waves results in efficient energy exchange between ions and
electrons so that the energy equilibration occurs in the upstream
flow before it arrives at the shock front. An important point is
that the energy of the maser radiation is generally proportional
to the energy of electrons entering the shock. Therefore
eventually a few percent of the total energy of the flow is
radiated away in the form of low-frequency waves.

Non-linear interactions of these waves with the plasma particles
could significantly affect the process of particle acceleration.
This issue demands multi-dimensional study and therefore is beyond
the scope of the present work. In any case, electron-ion
equilibration facilitates any acceleration process.

If the magnetization of the upstream flow is low,
$\sigma<10^{-2}$, the energy density of waves downstream of the
shock exceeds that of the shock-compressed background field.
Relativistic particles could radiate in the field of the waves via
nonlinear Compton scattering (e.g., Melrose 1980, pp. 136-141).
The power and characteristic frequencies of this emission are
similar to those for synchrotron emission in the magnetic field of
the same strength as the wave amplitude. Therefore the overall
spectrum of radiation from electrons with a power-law energy
distribution generally mimics the synchrotron spectrum. However,
there are marked differences. For example in the low-frequency
range, $\omega\ll (eE_0/m_ec)\gamma^2$, the radiation spectrum of
the nonlinear Compton scattering exhibits a larger slope than the
synchrotron spectrum, $F_{\omega}\propto \omega$ (He et al. 2003)
instead of the customary $F_{\omega}\propto \omega^{1/3}$.

Note that a good fraction of the gamma-ray bursts (GRBs) exhibit
X-ray spectra growing faster than $\omega^{1/3}$ (Preece et al.
1998). There is still no conventional explanation of this fact
(Lloyd \& Petrosian 2000; Medvedev 2000, 2006; M\'esz\'aros \&
Rees 2000; Fleishman 2006) The presented model could naturally
account for this behavior if one attributes the GRB emission, at
least partially, to the nonlinear Compton scattering of
low-frequency waves generated by the synchrotron maser at the
shock front.  Note also that observations of GRB afterglows
suggest (Waxman 1997; Frail, Waxman \& Kulkarni 2000; Eichler \&
Waxman 2005) that the electrons gain a significant fraction of the
total energy such that their average energy is comparable with
that of the ions. The present study shows that electrons indeed
take about one-half of the total energy, which makes it relatively
easy for any acceleration process to transfer a significant
fraction of this energy to a nonthermal tail.

I am grateful to David Eichler and Anatoly Spitkovsky for valuable
discussions. This research was supported by the grant
I-804-218.7/2003 from the German-Israeli Foundation for Scientific
Research and Development.

\section*{Appendix. Synchrotron instability}

Let us assume that all electrons rotate in the magnetic field with
the Lorentz factor $\Gamma\gg 1$. Such a plasma is evidently
unstable with respect to maser synchrotron emission. The growth
rate of the instability is maximal for the electromagnetic wave
propagating perpendicularly to the magnetic field and polarized
perpendicularly to the magnetic field; only this wave is
considered here.
\subsection{$\Omega_B\gg\Omega_p$}
In this case one can neglect the influence of the plasma on the
dispersion properties of the emitted waves. Then the growth rate
of the synchrotron instability at the $j$-th harmonics is easily
found as (e.g. Alexandrov, Bogdankevich \& Ruhadze 1984, sect.6.4)
$$
\kappa=\frac{\sqrt{3}}2\Omega_B\left[\frac{\Omega_p^2}{2\Omega_B^2j^2}J'^2_j(jV)\right]^{1/3};
$$
where $J_j$ is the Bessel function, $V=1-1/2\Gamma^2$ the electron
velocity. The growth rate is maximal at the first harmonics,
$\omega=\Omega_B=\Omega_p\sigma_e^{1/2}$, and may be written as
$$
\kappa_{max}=0.3\sigma_e^{-1/3}\Omega_B;\qquad \sigma_e\gg 1.
\eqno(A1)
$$

\subsection{$\Omega_B\ll\Omega_p$}
In this case only high harmonics of the gyrofrequency can
propagate, and therefore be emitted, within the plasma. Then the
maser instability develops only in the Rasin-Tsytovich range
unless the electron ring is unrealistically narrow (McCray 1966;
Zheleznyakov 1967). Sazonov (1970) and Sagiv \& Waxman (2002)
considered the maser instability of relativistic electrons with
isotropic distribution in the case where the dispersive properties
of the medium are determined by these electrons. In this section,
a similar problem is solved for a ring-like distribution.

Making use of the Einstein coefficient method (e.g. Ginzburg 1989,
ch.10) one can write the growth rate of the instability as the
absorption coefficient with the opposite sign:
$$
\kappa=\frac{8\pi^3c^2}{\omega^2}\sum A_i^j\left(N_i-N_j\right).
$$
Here $N_i$ is the population of the $i$-th Landau level.
Introducing the distribution function, $f(p_{\bot},p_{\|})$,
normalized by the condition $\int fd\mathbf{p}=1$ and making use
of the energy conservation in the emission process, one can write
$$
N_i-N_j=N\frac{\partial f(p_{\bot},p_{\|})}{\partial
p_{\bot}}\frac{\varepsilon}{p_{\bot}}\hbar\omega;
$$
where $\varepsilon$ is the electron energy, and $p_{\bot}$,
$p_{\|}$ are the components of the momentum perpendicular and
parallel to the magnetic field. The Einstein coefficients,
$A_i^j$, are straightforwardly expressed via the conventional
formula for the power of the synchrotron radiation from a single
electron, $q$; this yields
$$
\kappa=\frac{4\pi^2 cN}{\omega^2}\int q\frac{\partial
f(p_{\bot},0)}{\partial p_{\bot}}p_{\bot}^2dp_{\bot};\eqno(A2)
$$$$
q=\frac{3^{1/2}e^2\omega_B}{2c[1+\gamma^2(1-n^2)]^{1/2}}\frac{\omega}{\omega_c}
\left[\int_{\omega/\omega_c}^{\infty}K_{5/3}(z)dz+K_{2/3}\left(\frac{\omega}{\omega_c}\right)\right];
$$$$
 \omega_c=\frac 32\omega_B\gamma^2[1+\gamma^2(1-n^2)]^{-3/2};
$$
where $K_j$ is the Macdonald function,  $n$ the refraction index
of the medium and it was assumed that $\gamma\gg 1$ so that the
electron radiates exactly in the direction of its motion.
Integrating equation (A2) by parts, one gets
$$
\kappa=\frac{\pi\omega_p^2m_e}{\sqrt{3}\omega}\int
f(p_{\bot},0)\frac{\partial}{\partial
p_{\bot}}\left\{p_{\bot}^2\left(n^2-1-\frac
1{\gamma^2}\right)\left[\int_{\omega/\omega_c}^{\infty}K_{5/3}(z)dz+K_{2/3}\left(\frac{\omega}{\omega_c}\right)\right]\right\}
dp_{\bot}\eqno(A3)
$$

The refraction index of the wave propagating transversely to the
magnetic field and polarized transversely to the magnetic field is
expressed via the transverse dielectric permittivity as
$n=\sqrt{\epsilon}$, where (e.g. Alexandrov et al. 1984)
$$
\epsilon=1+\frac{\omega_p^2}{\omega}\int\frac{p_y}{\gamma(\omega-kv_x)}
\frac{\partial f}{\partial p_y}d\mathbf{p}.
$$
For a narrow relativistic ring with the distribution function
$$
f(p_{\bot},p_{\|})=\frac 1{4\pi^2m_e^3\Gamma(\delta\Gamma)^2}
\exp\left(-\frac{(p_{\bot}-m_e\Gamma)^2+p_{\|}^2}{2(\delta\Gamma)^2}\right)
;\eqno(A4)
$$
where $\delta\Gamma\ll\Gamma$, one finds
$$
\epsilon=1+\frac{\Omega_p^2}{k^2}\left\{1-\sqrt{1-\frac{k^2V^2}{\omega^2}}
-\frac{k^2V^2}{\omega\Gamma^2\sqrt{\omega^2-k^2V^2}}\right\}.
$$
In the case $\omega\gg\Omega_p$ so that $n-1\ll 1$, the refraction
index takes the usual form
$$
n^2=1-\frac{\Omega_p^2}{\omega^2}.
$$
In the Rasin-Tsytovich regime, $\Gamma^2(n^2-1)\gg 1$, equation
(A3) yields for the distribution function (A4)
$$
\kappa=\frac{\Omega_p^4\Gamma}{\sqrt{6\pi}\omega^3\delta\Gamma}
F\left(\frac{2\Omega_p^3}{3\Omega_B\omega^2}\right);
$$
where
$$
F(z)=zK_{5/3}(z)-\int_z^{\infty}K_{5/3}(x)dx-(4/3)K_{2/3}(z).
$$
The function $F(z)$ achieves its maximum at $z=2.2$ and its
maximal value is 0.085. Now one can estimate the maximal growth
rate of the instability as
$$
\kappa_{max}=0.1\frac{\Gamma}{\delta\Gamma}\Omega_B\sigma_e^{1/4};
\qquad \sigma_e\ll 1. \eqno(A5)
$$
The corresponding emission frequency is
$\omega=0.5\Omega_p\sigma_e^{-1/4}$.

\includegraphics[width=10 cm,scale=1.2]{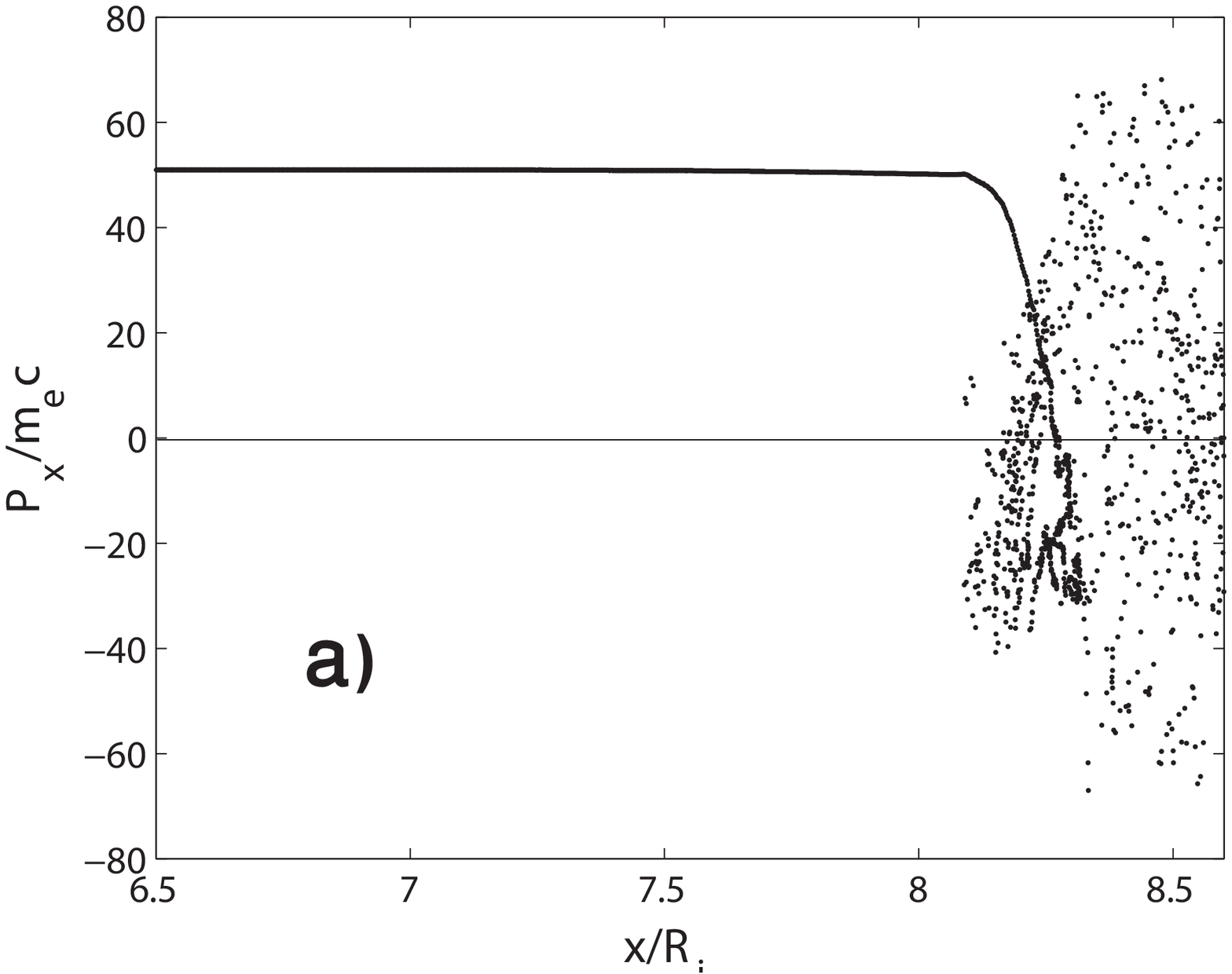}

\includegraphics[width=10 cm,scale=1.2]{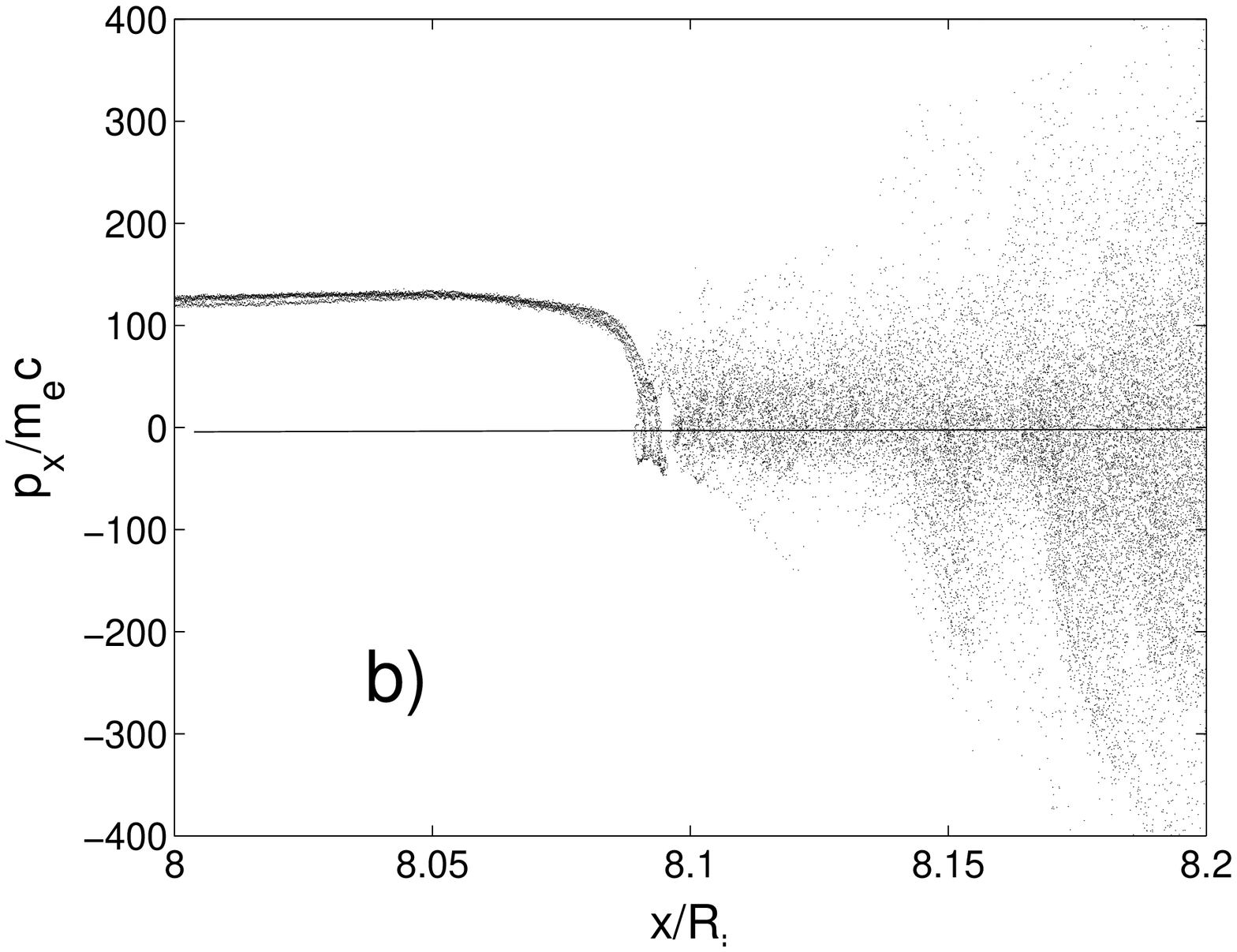}

\includegraphics[width=10 cm,scale=1.2]{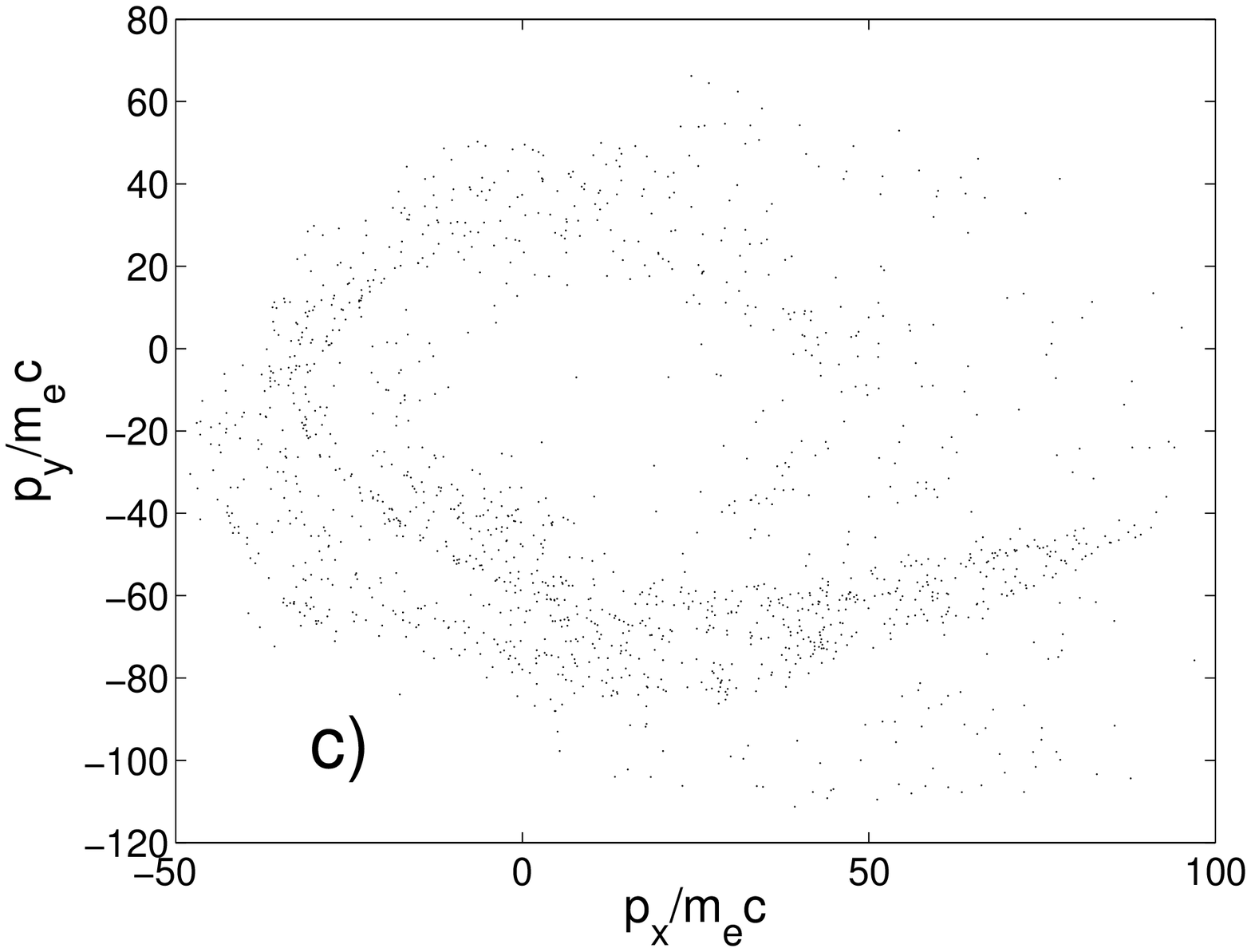}

\includegraphics[width=10 cm,scale=1.2]{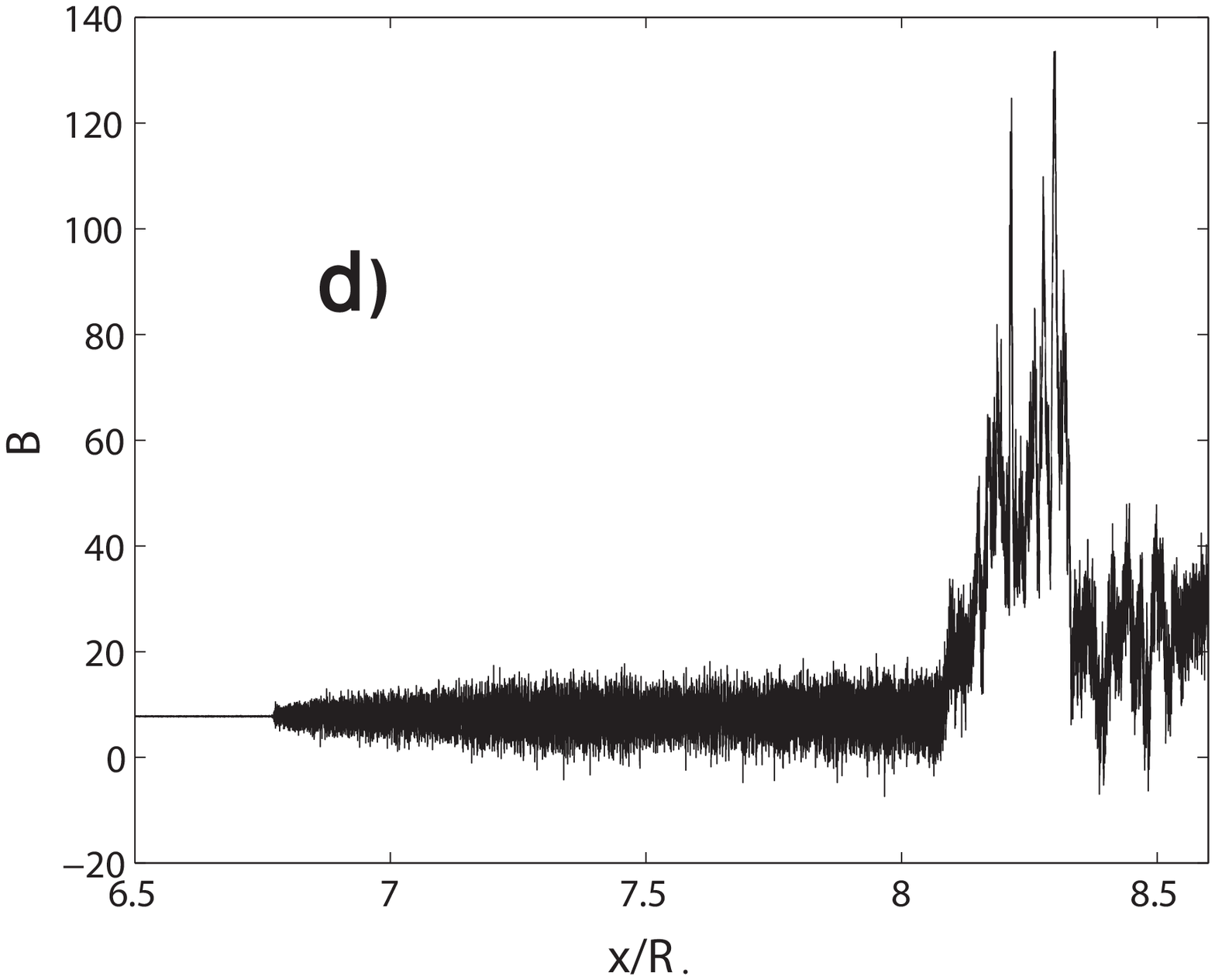}

\begin{figure*}
\includegraphics[width=10 cm,scale=1.2]{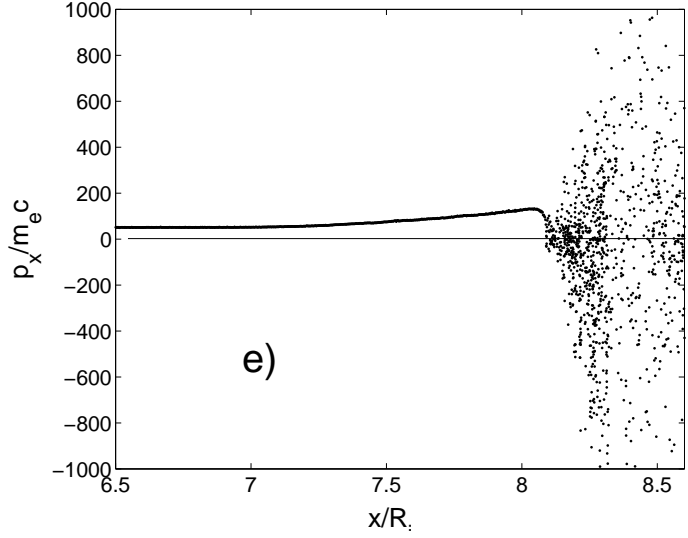}
\caption{Electron and ion phase space and electromagnetic fields
at $t=1.9R_i$; the initial Lorentz factor $\Gamma_0=50$, the
magnetization $\sigma_i=0.003$, the ion-electron mass ratio
$M/m=200$. a) longitudinal momentum of ions; b) longitudinal
momentum of electrons at the front of the shock; c) electron phase
space at the front of the shock, $8.088<x/R_i<8.101$; d) magnetic
field; e) longitudinal momentum of electrons}
\end{figure*}

\begin{figure*}
\includegraphics[width=10 cm,scale=1.2]{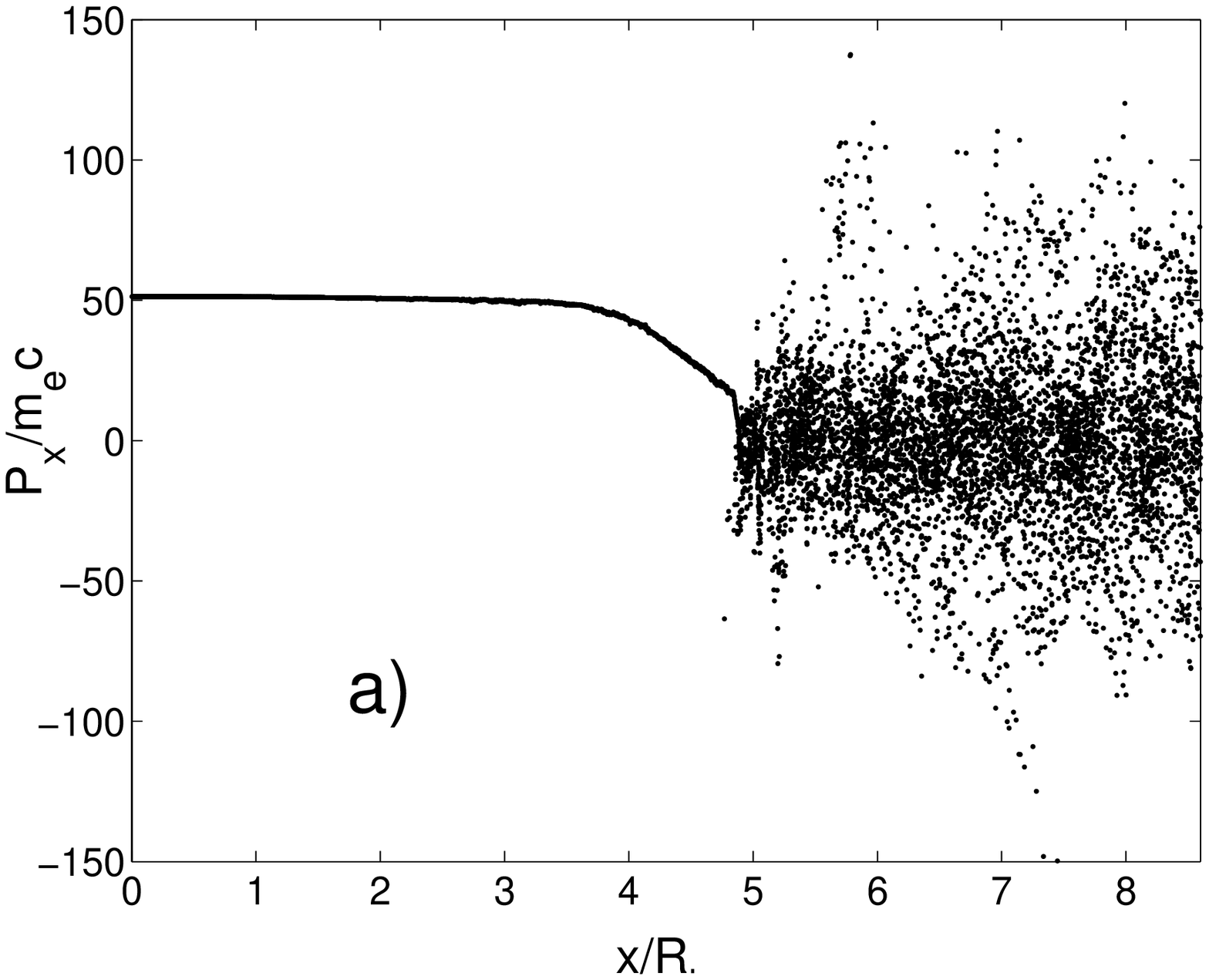}
\end{figure*}
\begin{figure*}
\includegraphics[width=10 cm,scale=1.2]{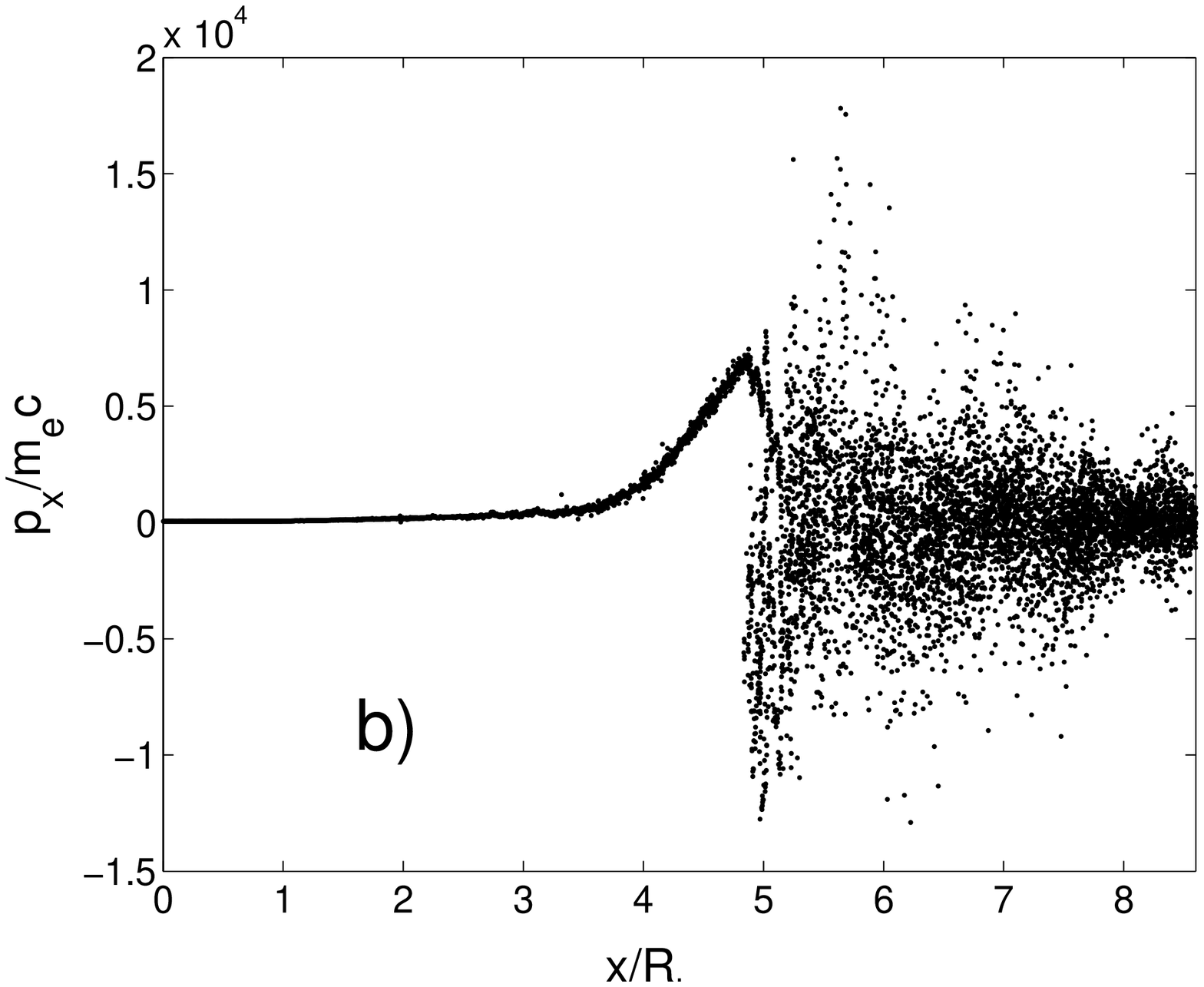}
\end{figure*}
\begin{figure*}
\includegraphics[width=10 cm,scale=1.2]{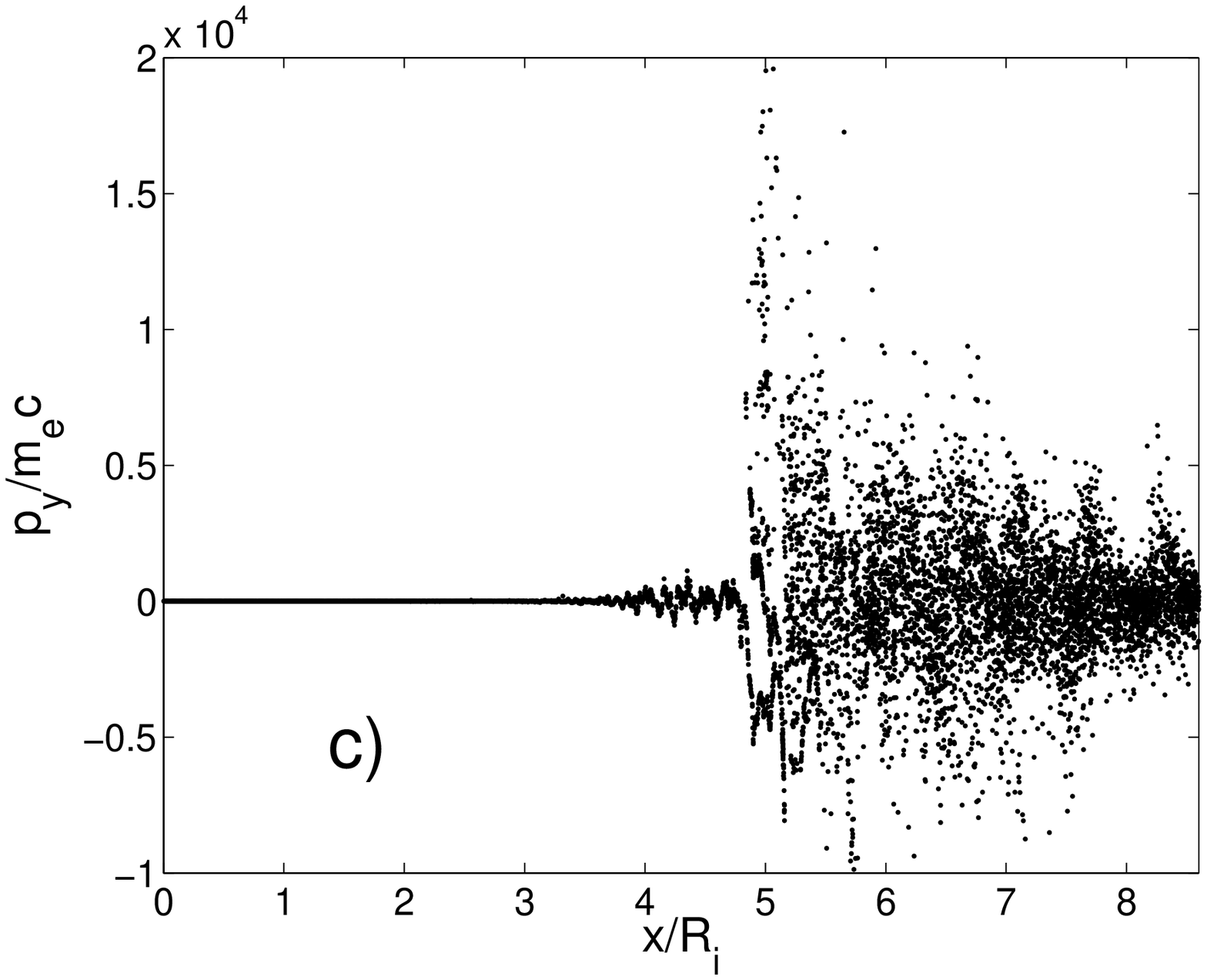}
\end{figure*}
\begin{figure*}
\includegraphics[width=10 cm,scale=1.2]{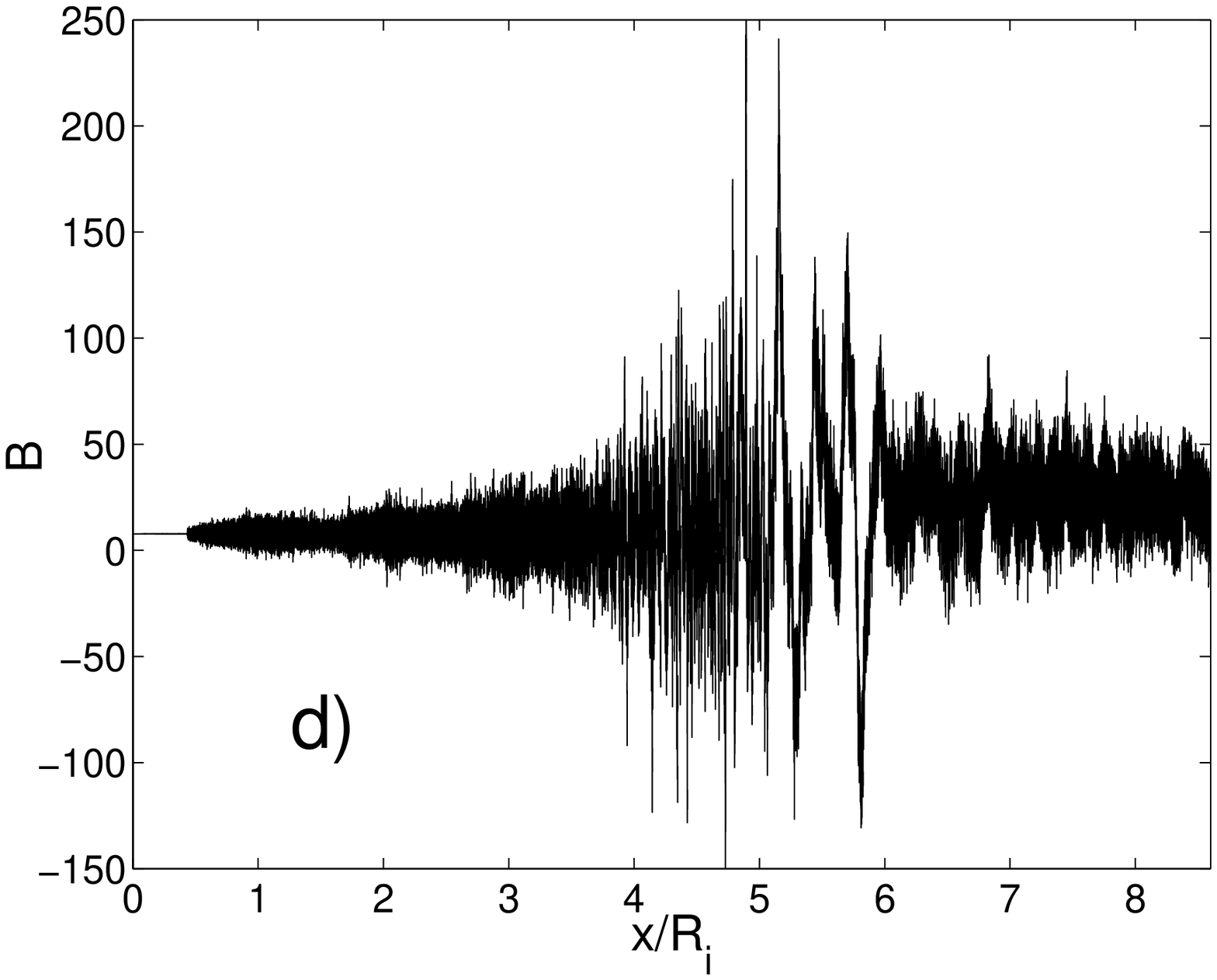}
\end{figure*}
\begin{figure*}
\includegraphics[width=10 cm,scale=1.2]{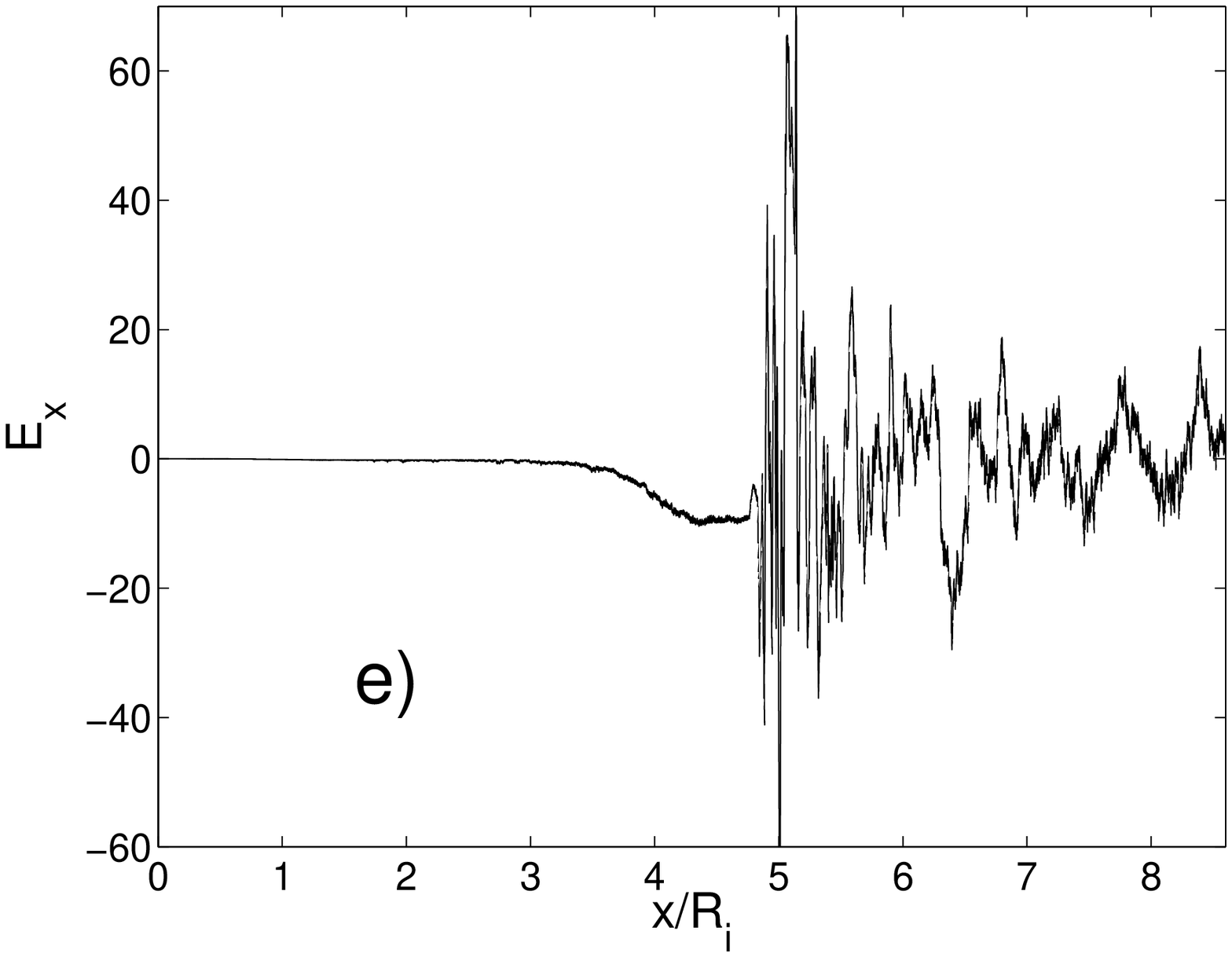}
\caption{Electron and ion phase space and electromagnetic fields
at $t=8.2R_i$. Parameters are the same as in Fig.1. a)
longitudinal momentum of ions; b) longitudinal momentum of
electrons; c) transverse momentum of electrons; d) magnetic field;
e) longitudinal electric field}
\end{figure*}
\end{document}